\documentclass[letterpaper]{article} 
\usepackage{aaai25}  
\usepackage{times}  
\usepackage{helvet}  
\usepackage{courier}  
\usepackage[hyphens]{url}  
\usepackage{graphicx} 
\usepackage{natbib}  
\usepackage{caption} 
\usepackage{booktabs} 
\usepackage{algorithm}
\usepackage{algorithmic}
\usepackage{enumitem}
\usepackage{comment}

\usepackage{newfloat}
\usepackage{listings}
\DeclareCaptionStyle{ruled}{labelfont=normalfont,labelsep=colon,strut=off} 
\floatstyle{ruled}
\newfloat{listing}{tb}{lst}{}
\floatname{listing}{Listing}
\title{}
\author{
    Written by AAAI Press Staff\textsuperscript{\rm 1}\thanks{With help from the AAAI Publications Committee.}\\
    AAAI Style Contributions by Pater Patel Schneider,
    Sunil Issar,\\
    J. Scott Penberthy,
    George Ferguson,
    Hans Guesgen,
    Francisco Cruz\equalcontrib,
    Marc Pujol-Gonzalez\equalcontrib
}

\title{Advancing Data Equity: Practitioner Responsibility and Accountability in NLP Data Practices}
\author {
    Jay L. Cunningham\textsuperscript{\rm 1},
    Kevin Zhongyang Shao\textsuperscript{\rm 2},
    Rock Yuren Pang\textsuperscript{\rm 2},
    Nathaniel Mengist\textsuperscript{\rm 2}
}
\affiliations {
    \textsuperscript{\rm 1}DePaul University\\
    Chicago, IL USA\\
    \textsuperscript{\rm 2}University of Washington\\
    Seattle, WA USA\\
    jcunni37@depaul.edu, kshao918@uw.edu, ypang2@uw.edu, mengin@uw.edu
}

\begin{document}

\maketitle

\begin{abstract}

While research has focused on surfacing and auditing algorithmic bias to ensure equitable AI development, less is known about how NLP practitioners, those directly involved in dataset development, annotation, and deployment, perceive and navigate issues of NLP data equity. This study is among the first to center practitioners’ perspectives, linking their experiences to AI governance frameworks and advancing participatory recommendations that bridge technical, policy, and community domains. Drawing on a 2024 questionnaire and focus group, we examine how U.S. based NLP data practitioners conceptualize fairness, contend with organizational and systemic constraints, and engage emerging governance efforts such as the U.S. AI Bill of Rights. Findings reveal persistent tensions between commercial objectives and equity commitments, alongside calls for more participatory and accountable data workflows. We critically engage debates on data diversity and ``diversity-washing,'' arguing that improving NLP equity requires structural governance reforms that support practitioner agency and community consent.
\end{abstract}

\section{Introduction}
As language technologies permeate every facet of society, the need for responsible and inclusive dataset practices has never been more critical. Recent scholarship has documented persistent challenges in natural language processing (NLP) systems, including automated speech recognition (ASR), text classification, and language generation \cite{koenecke_racial_2020, sap_risk_2019, deas_evaluation_2023}. Technologies such as smart virtual assistants and large language models often perform unfairly, reflecting biases rooted in dataset production practices, including limited diversity in training data and in the technologists who develop these systems \cite{bender-friedman-2018-data, goyal_is_2022}. In data science for machine learning (ML), dataset production involves multiple stages—curation, collection, annotation, and implementation \cite{muller2019data}; each shaped by human value judgments that can encode implicit biases toward or against certain language varieties.

These practices can have serious real-world consequences. Recent cases include low-income older Americans being denied health benefits because of algorithmic decisions and a pregnant Black American woman wrongfully arrested due to flawed facial recognition technology. In response, fairness research communities within CHI, FAccT, and AIES have emphasized the need for direct participation by underrepresented socio-cultural groups in NLP data production \cite{denton_whose_2021}. Community-driven initiatives such as DataWorks have demonstrated participatory approaches by annotating African American English (AAE) datasets to address representational harms~\cite{rothschild2022interrogating}. Policymakers have also advanced AI governance frameworks, including the U.S. AI Bill of Rights, to safeguard against socio-technical harms and promote socially responsible design \cite{lazar2023ai}.

While prior work has examined fairness across AI/ML pipelines \cite{holstein2019improving, miceli_between_2020, thakkar2022machine, widder2023dislocated}, NLP presents distinct equity challenges that are comparatively underexplored. The rapid deployment of large language models has transformed practitioners’ roles, intensifying pressures to address equity within complex organizational contexts. Yet there remains limited empirical understanding of how NLP practitioners themselves perceive and navigate data equity challenges, despite their direct influence on dataset integrity and, ultimately, system behaviors \cite{gebru_datasheets_2021, suchman_located_2002}.

We define NLP data equity as the practice of ensuring that NLP datasets are representative, contextually accurate, and co-produced in ways that acknowledge and mitigate structural inequities affecting marginalized groups. This requires both diversifying data sources and addressing power asymmetries in collection, annotation, and model deployment to support fair outcomes and epistemic justice \citep{barocas2020fairness, hanna_towards_2020, denton_whose_2021}.

Our study includes practitioners working with both text-based and speech-based NLP systems, including projects such as ASR, virtual assistants, sentiment analysis, machine translation, and language generation\footnote{(See the Appendices (C) in extended paper version)}. We therefore treat data equity as a cross-cutting concern across modalities. Where relevant, we note that certain issues—such as linguistic bias, representational harm, and dialect recognition—may manifest differently in speech versus text systems. Insights from this study may inform both domains.

This study investigates how NLP practitioners engage with data equity within their professional contexts. We focus on their awareness, decision-making processes, and perceived constraints, as well as their perspectives on governance and policy. \\
\textbf{We ask:}
\begin{enumerate}
\item How do NLP practitioners understand and navigate NLP data equity challenges within their workflows, and how are these shaped by organizational constraints and accountability structures?
\item What strategies, training, and resources do practitioners report using or lacking to identify and mitigate bias in NLP datasets?
\item How do practitioners perceive the role of governance, policy, and multi-stakeholder engagement in advancing equitable NLP data practices?
\end{enumerate}

While past work has examined practitioner awareness of fairness in AI/ML broadly (e.g., Holstein et al., 2019; Deng et al., 2023), few studies center on NLP-specific data practices and practitioners’ agency in navigating fairness-related constraints in industry and research settings. Our study bridges policy and practice by examining how practitioners perceive and implement AI governance frameworks.

\textit{*An extended version of this paper, which includes supplementary materials, technical appendices, and expanded methodological details, is available on arXiv\footnote{navigate to \url{https://jaylcunningham.com/work} for the arXiv web-link.} under a Creative Commons Attribution (CC BY) license.}

\section{Related Work}
\subsection{Critical Approaches in ML Data Practices}
Critical approaches have been interrogating the foundational assumptions that shape the ML dataset production, and advocating for strategies to reduce bias and ensure the inclusion of socio-culturally diverse data in ML applications \cite{mothilal2024towards}. For instance, \emph{Data Feminism} calls for a power redistribution in data practices to achieve greater equity. \citet{jo_lessons_2020} highlights the limitations in many existing datasets, proposing more responsible data practices that prioritize consent, inclusion, transparency, power dynamics, and privacy. Additional recommendations included assembling diverse pools of annotators, incorporating iterative feedback processes, and fine-tuning models to account for ethnic dialects, alongside improving documentation protocols (e.g., datasheets for datasets)~\citep{gebru_datasheets_2021, ziman2018automatic, kapania_hunt_2023, miceli_documenting_2022, pang_auditing_2023, raji_ai_2021}. While data equity has been extensively studied in AI/ML, nuances on NLP-specific issues, such as linguistic representation, dialect variance, and cultural annotation complexity, are often overlooked~\cite{deng2024promoting}. Moreover, the emergence of LLMs has changed how NLP practitioners interact with data curation workflows. 

Among these strategies, participatory design~\cite{muller1993participatory} has been used to improve ML systems by involving community stakeholders directly in the dataset production process, ensuring that the resulting data accurately represent diverse perspectives \citep{sloan2020participation, cunningham_collaboratively_2023}. Human-centered methods, such as community workshops, co-design sessions, and feedback mechanisms, are recommended as research tools to align with participatory modes of NLP dataset production. These methods promote a more inclusive approach to data creation by integrating input from the communities whose language and perspectives are represented in the data, thus contributing to more equitable and accurate models. Our work also intersects with HCI scholarship on data infrastructures and activism, which highlights how socio-technical participation is constrained and enabled by institutional, material, and political conditions \cite{irani2016hackathons,dourish2017stuff}.

\subsection{Challenges of NLP Data Equity}
As we defined, NLP data equity refers to a practice ensuring that NLP datasets themselves are representative and contextually accurate, as well as co-produced in ways that acknowledge and mitigate inequities affecting marginalized groups. However, achieving this ideal faces challenges, particularly given the rapid pace of AI development that often prioritizes speed over equitable data practices.
For example, \citet{paullada2021data} delineates the limitations in historical and contemporary NLP data practices, focusing on the challenges of dataset design, development, and introspection, alongside the broader practices surrounding dataset usage. In particular, tensions between the NLP system performance and the data of linguistically diverse users expose the mis-alignments between the creation of the dataset and the real-world linguistic diversity, presenting complex socio-technical challenges that require critical reflection on current data practices \citep{mathur2022we, joshi2020state}. 

Implicit biases perpetuated in core NLP data practices such as transcription and annotation have been identified as key contributors to inadequate AI models, potentially exacerbating societal inequalities \citep{hutchinson-etal-2020-social}. One illustrative example involves the Perspective API, which assigned disproportionately high toxicity scores to tweets featuring vernacular elements and linguistic structures associated with African American English (AAE) \citep{sap2019risk}. Such discrepancies highlight the inadequacies of models that fail to account for sociolinguistic diversity, often leading to suboptimal performance for users who communicate in minority language varieties \citep{kamikubo_data_2022}. This challenge is particularly pronounced for speakers of AAE, where existing NLP models have been shown to inadequately due to insufficient linguistic representation \citep{tatman_gender_2017, cunningham2024understanding}.

Scholarly discourse has also identified a misalignment between industry metrics for successful performance and the realities of socially situated, culturally nuanced human communication. For instance, \citet{kapania_hunt_2023} identified primary obstacles to incorporating diversity into data practices: insufficient information about annotators, operational fragmentation, and the competing priorities in ML development. To address these issues, some researchers propose a participatory framework in which practitioners collaborate with annotators familiar with the linguistic varieties \citep{kapania_hunt_2023, sloane_participation_nodate}. 

Our study builds on this body of work by empirically exploring the perspectives of NLP practitioners regarding these challenges, while further identifying the potential for collaborative co-creation of datasets \citep{kapania_hunt_2023, paullada2021data, deas_evaluation_2023}. Additionally, it expands upon previous research by examining the integration of stakeholder participation across multiple stages of data practices, beyond the scope of annotation alone \citep{kapania_hunt_2023, mcconvey2024not}. To this regard, we integrate critiques of synthetic diversity and diversity-washing \cite{whitney2024diversitywashing, dalal2024provocation}, highlighting how superficial diversity initiatives can obscure systemic inequities. Ultimately, this study builds on prior work \cite{holstein2019improving, 10.1145/3593013.3594037} by focusing specifically on how NLP practitioners interpret ML fairness frameworks, regulatory guidelines, and equity mandates in real-world contexts.

\subsection{AI Policies on Addressing Algorithmic Disparities} 
The introduction of policies and legislation has emerged as a response to the growing societal challenges posed by NLP technologies, as exemplified by initiatives such as the EU AI Act \cite{act2024eu}, the U.S. White House AI Bill of Rights \cite{U.S.WhiteHouse_AIBillofRights_2022}. 
In 2023, the National AI Research Resource Task Force released a report providing recommendations for the establishment of a public resource to support AI research in the United States, emphasizing the need to consider societal impacts during the review of proposals that seek access to public resources. The U.S. AI Bill of Rights further emphasizes that AI-driven automated systems must actively and continuously protect the public from algorithmic discrimination. It stipulates that data used in system development or evaluation must be representative of the local communities where the AI will be deployed, and that this data should be scrutinized for bias within its historical and societal context to identify and mitigate potential harms.

Although these policies largely concentrate on downstream applications of AI systems, questions remain regarding the practical implementation of the proposed measures by technologists. In particular, NLP training datasets, which are central to the accuracy of downstream tasks, are shaped by data science practices that play a critical role in shaping model outcomes. This study explores how NLP practitioners perceive the applicability of the U.S. AI Bill of Rights' guidelines in promoting diversity and safeguarding against algorithmic harms within their own work. The findings contribute to a broader understanding of the real-world challenges associated with applying policy frameworks to technical data workflows in the development of AI systems.

\subsection{AI Governance and Practitioner Responsibility}
AI governance has emerged as a critical area of scholarly and policy attention, encompassing the mechanisms, institutions, and practices through which the development, deployment, and oversight of AI systems are regulated \cite{floridi2018ai4people}. While much attention has focused on technical standards, policy instruments, and high-level regulatory frameworks \cite{jobin2019global}, emerging scholarship highlights the need for multi-scalar governance strategies that engage actors across the AI development life-cycle, including practitioners, communities, industry stakeholders, and regulators \cite{drahos2017regulatory}.

In the context of NLP systems, however, practitioner responsibility remains under-theorized, with limited attention to how data workers, engineers, and analysts are positioned within governance debates. Recent critiques have also called for expanding governance beyond compliance frameworks, addressing how inclusion, diversity, and fairness discourses in AI can risk reinforcing extractive or colonial data practices if not critically interrogated \cite{whitney2024diversitywashing, dalal2024provocation}. This critique is particularly salient for NLP practitioners navigating data curation and annotation tasks, where decisions about dataset composition, representational adequacy, and community engagement carry significant governance implications.

In parallel, recent scholarship has raised concerns about the limits of current governance approaches and diversity initiatives in AI development. While many efforts emphasize fairness principles or public-private partnerships, regulatory scholars argue for more expansive, multi-scalar governance frameworks that engage actors across policy, institutions, and the development pipeline \cite{drahos2017regulatory,jobin2019global}. These perspectives foreground the need to embed structural accountability and participatory oversight within AI systems development. Relatedly, emerging critiques of “diversity-washing” caution that efforts to increase dataset representativeness can obscure underlying power asymmetries or bypass community consent, particularly when synthetic or compliance-driven diversity metrics are used \cite{whitney2024diversitywashing,dalal2024provocation}. Our work builds on these insights by focusing on how NLP practitioners themselves perceive their roles within these governance and equity tensions—highlighting both the constraints they face and the opportunities for more accountable, community-centered data practices.

\section{Methods}
This study examines how some NLP practitioners advance data equity by reflecting on practices that produce representative, contextually accurate, and justice-oriented datasets to address technical misalignments and social inequities among diverse users.
The analysis focuses on practitioner perspectives and experiences, rather than direct audits of NLP systems, to understand how organizational contexts, governance frameworks, and individual agency shape equity-oriented data practices.

We used a mixed-methods design, including a survey ($N=47$) and a focus group ($N=5$), to capture both breadth and depth in practitioner perspectives. Surveys have been effective in mapping practitioner attitudes in NLP fairness research \citep{sambasivan_re-imagining_2021}, while focus groups enable richer, context-specific insights \citep{miceli_documenting_2022}. Our single focus group used purposive sampling of “critical practitioners” to probe reasoning processes behind survey trends; this approach prioritizes transferability over statistical generalization \citep{lincoln1985naturalistic}.

Survey responses were analyzed descriptively in Microsoft PowerBI using frequency distributions, cross-tabulations, and percentages. Open-ended responses and focus group transcripts were coded thematically to identify patterns in bias mitigation experience, governance engagement, and organizational constraints.

All study procedures were reviewed and approved by the University of Washington Institutional Review Board (IRB Protocol). 

\subsection{Questionnaire}
The goal of the questionnaire was to understand on current NLP data practices in a lightweight and accessible manner. Specifically, the questionnaire was used to understand the strategies employed by NLP professionals to mitigate bias and ensure representation within datasets. Given the semi-structured, exploratory nature of the study, some questions were informed by topics of related works and developed through unvalidated measures \cite{hill_educators_2022} to address our research questions. The questionnaire was administered via the online platform (Typeform) in late 2023, and responses from 47 eligible participants were analyzed. 

\textbf{Questionnaire Recruitment and Respondents}
The study targeted self-identified NLP data practitioners with professional experience in text and speech NLP systems, encompassing roles such as data engineers, data scientists, data annotators, data researchers, and crowdworkers. Recruitment strategies included outreach via online social networking platforms (e.g., X (Twitter), LinkedIn, Reddit, GroupMe), direct emails, and mailing lists from both internal and external organizations. Additionally, professional networks based on prior collaborations with academic and industry AI teams (e.g., Google, Microsoft, Apple, and Meta) were leveraged. The questionnaire started with an information statement detailing the objectives, benefits, and incentives of participation, followed by a disclosure of study procedures and an acknowledgment of informed consent. Eligibility screener questions ensured participants were at least 18 years old, resided and worked in the United States, and had professional experience in NLP dataset production.

\textbf{Questionnaire Design}
The questionnaire comprised a total of 27 questions\footnote{(See the Appendices (A) in extended paper version)}, categorized into quantitative (i.e., 11 multiple-choice, 5 multi-select, 3 rating scale questions) and qualitative: (i.e., 8 open-ended questions). Subsequently, the questionnaire explored a variety of themes that were mapped to the study's research questions. The questionnaire concluded with questions on demographic information and their interest in engaging in subsequent focus group studies.

\textbf{Participant Demographic Details}
We collect and report context regarding demographics of study participants as a way to evaluate claims of data practitioner identity on the quality and performance of ML systems. Out of the 60 respondents who began the questionnaire, 13 were deemed ineligible or declined to complete the survey, resulting in 47 eligible responses for analysis. These respondents, with an average age of 29 (ranging from 19 to 40), held various professional roles in NLP data practices, including researcher, scientist, and engineer positions. The majority of them (48.94\%) have an intermediate level of expertise for NLP data practices. Beginners account for 10.64\%, Advanced participants represent 25.53\%, and Experts make up 14.89\% of the surveyed group. Demographic data collected included gender identity, race and ethnicity, languages spoken, and educational credentials \footnote{(See the Appendices (C) in extended paper version)}. Specifically, the sample comprised 32 men, 13 women, one non-binary/non-conforming individual, and one participant who preferred not to disclose their gender identity. For those interested in participating in a drawing for a 50 USD Amazon gift card, email addresses were collected and aggregated into a participant interest pool. Participant demographic details are provided in (Table \ref{table 1}).

\begin{table}[!ht]
\centering
\small
\begin{tabular}{cc}
    \begin{tabular}[t]{ll}
    \toprule
    \textbf{Race/Ethnicity} & \textbf{\# of respondents} \\
    \midrule
    Black or African American & 20 \\
    \midrule
    White or Caucasian & 13 \\
    \midrule
    Asian & 10 \\
    \midrule
    American Indian or Alaskan Native & 1 \\
    \midrule
    Hispanic or Latin Origin & 1 \\
    \midrule
    Other & 2 \\
    \bottomrule
    \end{tabular}
\end{tabular}
\caption{Survey participant demographic distribution ($N = 47$) by race/ethnicity, highlighting the sample diversity.}
\label{table 1}
\end{table}
\subsection{Focus Group}
This study included a single focus group of five critically engaged NLP practitioners. While this limits the breadth of perspectives represented, the focus group was designed as an exploratory complement to the broader survey, offering depth rather than statistical saturation. Our intention was not to reach thematic saturation across the field but to capture detailed narratives that could illuminate the organizational and governance contexts shaping equity-related decision-making. The insights are thus transferable to similar professional and institutional settings, providing a foundation for future, larger-scale qualitative research.

\textbf{Focus Group Moderation}
In March 2024, two members of the research team facilitated a data analysis focus group involving five practitioners to extend the investigation and data analysis beyond the initial survey results. The focus group aimed to garner more nuanced qualitative data through insights gained from these practitioners supplementary to data from the initial questionnaire. By moderating an expert group discussion, we aimed to corroborate questionnaire insights and involve participants in data analysis through collective dialogue and consensus, further affirming informed strategies for enhancing fairness and broadening representation within NLP dataset production. The lead researcher directed participants to their provided guidebook
, then transitioned the group into the questionnaire insights review activity.  Participants were initially given  10-minutes to navigate through the PowerBI report
on their own and were asked  to review the report and jot notes of any interesting insights or analytics, in addition to any questions that they may have about the data to be discussed further.

\textbf{Recruitment and Participants} The inclusion criteria for the focus group mirrored those employed for the survey, ensuring consistency in participant selection. Out of 35 respondents who expressed interest in a 1.5-hour virtual expert focus group, five participants were recruited based on the subjective quality of their survey responses, including factors such as compelling content, evident interest in the topic, and nuanced contextual insights. The professional backgrounds of these participants were diverse, with two engaged in academic research and others specializing in data analysis, language engineering, and economic planning. Their areas of expertise spanned responsible AI, biotechnology, and economic planning. Each focus group participant received a \$75 USD gift card as an incentive for their successful participation. The focus group session was audio recorded using Jitsi, a free, open-source, and encrypted online conferencing software. Participants were instructed to disable their video cameras and avoid using profile images during the session to ensure anonymity. In contrast, the facilitating researchers' video feeds remained active to foster trust and confidence among participants. To further safeguard the study's integrity and participant privacy, each participant was assigned a code name (Table \ref{table4}) based on their professional title (e.g., researcher 1, engineer 2, annotator 3) for use throughout the focus group discussion.

\begin{table*}[t]
\centering
\begin{tabular}{lllll}
 \toprule 
 \textbf{Codename} & \textbf{Gender Identity} & \textbf{Age} & \textbf{Job Title} & \textbf{Industry/Field} \\ 
 \midrule 
    Researcher 1 (R1) & He/Him & 25 & Academic Researcher & PhD research \\
    \midrule
    Researcher 2 (R2) & She/Her & 24 & Academic Researcher & Responsible AI/Tech, Education \\
    \midrule
    Data Analyst 1 (DA1) & He/Him & 29 & Biotech Data Analyst & Bio tech \\
    \midrule
    Data Engineer 1 (DE1) & She/Her & 29 & Language Engineer & Natural language processing \\
    \midrule
    Data Analyst 2 (DA2) & He/Him & 33 & Economic Planner & Economics Planning \\
 \bottomrule 
\end{tabular}
\caption{Profile of focus group participants.}
\label{table4}
\end{table*}
\subsection{Study Data Analysis}
Our research team analyzed and visualized the questionnaire data using Microsoft Power BI, summarizing the data through quantitative and qualitative measures. The quantitative descriptive analysis approach facilitated a better understanding of the questionnaire-data distribution and variability through calculations of means, percentages, and number of responses. For instance, for questions involving multiple choice, multi-select options, and rating scales, we employed calculations alongside data visualizations to illustrate the distributions. We performed inferential analysis to make interpretations or inferences about the population based on the survey data \citep{marshall2011introduction}.

For open-ended questions, we employed a grounded-theory approach to code responses and thematically analyze them \citep{braun_using_2006, olson2014ways}. We conducted multiple rounds of open coding involving multiple members of the research team to connect the broader qualitative excerpts with overarching themes of the survey study areas, thus informing the findings. To effectively communicate sentiments from participants, we include direct quotes from in the Findings section, with the survey respondents mentioned broadly as “respondent,” and focus group participants by their assigned participant ID (e.g., R1) to differentiate them in the analysis.

\section{Findings}

The findings presented here are not intended to be statistically generalizable across all NLP practitioners. Instead, they are analytically transferable to  contexts where practitioners operate under comparable organizational conditions and governance constraints. This framing aligns with qualitative research principles that emphasize the depth and contextual richness of insights over broad representativeness.

Survey findings establish broad patterns in practitioner experiences and perspectives, while focus group insights add depth from critically engaged participants. The survey data provides a wide view of how NLP professionals encounter and interpret data equity challenges, while the focus group offers rich, contextual accounts of equity motivations, organizational constraints, and governance engagement. Together, these methods illuminate both the prevalence of certain views across the field and the underlying structural and institutional factors that shape them.

In this section, survey questions are referenced using the notation “SQ” followed by the question number (e.g., SQ-10c). Focus group excerpts are labeled “FG” with a sequential number (e.g., FG-3). These labels correspond to survey or focus group questions. \footnote{(See the Appendices (B) in extended paper version)}

\subsection{Practitioners’ Applied Data Practices are Harms Neutral
}
The initial set of survey questions sought to understand the range of experiences of NLP practitioners in organizing, collecting, transcribing, and analyzing text or speech data for NLP datasets. The survey study starts with surfacing the various areas of experience among practitioners in \textbf{[SQ-6a]}. The analysis primarily indicated that the data work of most NLP practitioners was centered on data collection or annotation for training of the text / speech model ($76.60\%$), research on data practices or methods ($46.81\%$) and organizing and managing data annotation projects ($42.55\%$). These activities were particularly relevant to NLP technologies such as automated speech recognition (ASR), virtual assistants, smart home devices, and sentiment analysis. The full spectrum of NLP technologies that NLP practitioners reported working on. \footnote{(See the Appendices (C) (Table 3) in extended paper version)} However, we find that practitioners had the least experience mitigating bias in data training sets ($23.40\%$) and developing evaluation approaches to measure the effectiveness of training data (31.91\%).


This context indicates that while NLP practitioners are extensively engaged in the applied operational facets of data handling, they can be neutral to emergent harms, exposing deficiencies in practices related to recognizing and mitigating potential biases, along with  evaluating the outcomes of training datasets for users and systems. This discrepancy directed our focus toward understanding the underlying reasons for the limited engagement in harms reducing practices.

\subsubsection{Barriers to Bias Mitigation}
The focus group discussions provided deeper insight into why might NLP practitioners to be less experienced in mitigating bias \textbf{[FG-6a]}, revealing both practical and philosophical challenges that may hinder efforts in this area. Participant DE1 expressed concerns over the potential repercussions of identifying biases, including financial costs associated with mitigation strategies and negative impacts on organizational reputation, which may discourage proactive measures in bias mitigation. 
In addition, the mention of "short contracts don't meet basic needs" highlights another practical issue—temporary or short-term visions in NLP projects might not provide the stability or incentives needed for workers to engage deeply with complex issues like bias mitigation.

\subsection{Circumstances of Integrating Diverse  Data and Inclusion Conscious Practices}
Participants reflected on how the increasing reliance on LLMs reduces attention to equity-oriented data goals and raises the risk of flattening sociolinguistic nuance. Some voiced skepticism toward diversity as a compliance goal rather than a community-driven value. Our findings give insight into NLP practitioners’ efforts for achieving diversity and representation within data science practices. 9 out of 47 survey respondents elaborated on their efforts to strive for diverse representation and inclusion in their curated datasets\textbf{ [SQ-7a]}. Notably, 28 respondents indicated that striving for diverse variations in their data collections to improve users inclusion and system performance were either not applicable to their roles or were not considerable factors.

The findings from this analysis indicate that a marginal percent of NLP practitioners are actively taking steps to promote diversity and inclusion in their datasets. 
Four primary themes emerged from the responses in relation to circumstances contributing to NLP data integration:
\begin{itemize}
    \item \textbf{Seeking data from diverse sources: }
    Respondents emphasized their effort to acquire data from various sources representing different cultures, demographics, and communities, including online communities, minority-specific resources, and diverse media outlets. 
    \item \textbf{Ethical and responsible data efforts with diverse teams:} Respondents mentioned ethical and responsible data collection methods to ensure representativeness and reduce bias. They also highlighted the role of diverse teams in identifying and mitigating biases within datasets. 
    \item \textbf{Robust data quality and auditing practices} Ensuring data quality through regular reviews and updates was identified as essential for maintaining dataset relevance and reflecting evolving cultural nuances of language. Techniques such as pre-processing, feedback loops, and verifying data accuracy across different demographics were mentioned as methods to mitigate bias and ensure accurate representation of target populations.
    \item \textbf{Cultivating a technical culture that prioritizes inclusion:} Respondents highlighted the importance of prioritizing inclusive awareness with the data-centric product team, particularly when project priorities vary based on stakeholder and business interests. 
\end{itemize} 


\subsubsection{Limitations in Sourcing Diverse Datasets} 21 survey respondents mentioned challenges in sourcing diverse datasets. To further understand this, we explored \textbf{[FG-7a]} the limitations and obstacles faced by NLP practitioners in the focus group discussion. Several barriers were revealed including: 1) Data volume and representation (e.g., "not all populations might ... provide the same ’volume’ of data, to be considered across scale."), 2) Considerations for diverse data variations (e.g., "data should be diverse and cover all races and all areas, not one region only."), and 3) Cost and practical constraints (e.g., "a company will look at how they can produce the most data for the least cost first before considering ... diversity search.").

\subsubsection{Consideration of Social Demographics and Cultural Nuances} 
With 17 respondents indicating that social demographics and cultural nuances were not applicable to their roles in \textbf{[SQ-7b]}, the remaining 30 responses suggest that a growing number of NLP practitioners recognize its importance in ensuring diverse and representative datasets (e.g., "consider diverse social
demographics and nuances by ensuring a broad representation..."). 
In addition, a few specific aspects of social demographics and cultural nuances were intentionally taken into considerations by participants. These include: 1) Race/Ethnicity; 
2) Language: Accents, dialects, and English varieties; 
3) Cultural background and sensitivities;
4) Regional variations based on country, state, and geographic areas; 
5) Other aspects such as age, gender, income-level or socioeconomic status, accessibility challenges or disabilities (e.g., stuttering, stammering), and sexual-identity (LGBTQAI+). It was also noted that certain social identity aspects were more relevant to some NLP technologies than others. For example, one respondent noted “these features weren’t included when we applied sentimental analysis,” whereas another stated, “in my own work within ASR research, in considering speakers of AAVE, we consider ethnic, cultural, and linguistic aspects.” 

\subsection{Defining and Measuring Success in NLP Dataset Annotation}
Practitioners described challenges in annotating culturally relevant text, curating datasets for diverse dialects, and aligning with internal metrics. These issues are unique to NLP systems where meaning is deeply context-dependent.

\subsubsection{Assessing the Implications of Data Annotation Practices} Our survey showed the diverse criteria used by NLP practitioners to assess the performance of data annotators and their practices \textbf{[SQ-8a]}. A substantial majority of the respondents identified the quality of annotations ($68.09\%$) and the precision of annotations ($55.32\%$) as primary indicators of success, which underscores that producing reliable and precise annotations is essential for effective NLP model training. The respondents also indicated other applicable options: compliance with annotation guidelines ($44.68\%$), feedback from other annotators ($34.04\%$), and speed of annotation completion ($31.91\%$). 
Another two respondents mentioned alternative factors such as participation in ``internal annotator agreements'' and comprehensive data collection efforts (e.g., ``recruiting and paying participants, collecting their data/paperwork, and recording the tasks.'').

\textit{Quality of Annotation} and \textit{Accuracy of Annotations} were among the highest rated factors based on survey analysis, so our focus group discussions centered on how is quality and accuracy defined for data annotators and who develops metrics of success for data annotators \textbf{[FG-8a]}.
One key finding is that the notions of ``quality'' and ``accuracy'' in annotation are inherently subjective. While factors like adherence to guidelines, consistency, and the overall usability of the annotated data are commonly cited, there is no universal definition that applies across all projects. 

The focus group responses highlighted the tension between speed and quality as performance metrics. DE1 noted, ``I spent two-ish years as an annotator and speed was considered a much more important factor as meeting deadlines was more important than having more quality data...'' 
This underscores some projects prioritize meeting deadlines and high-volume data production, while others emphasize annotation precision. This tension varies with the project's scale, data type, and goals (R1). Approaches to develop metrics for annotator success are diverse. Some approaches involve collaborative input from annotators, while others rely on a top-down approach by managers or clients (DE1). Based on participant responses, clear and precise guidelines are universally valued for steering annotation efforts and defining success metrics.

Additionally, survey insights highlighted the significance of time which is typically allocated to handle conflicts of opinion among annotators and their processes \textbf{[SQ-8b]}. The survey indicated that 30 of 47 respondents allocated a moderate to substantial amount of time to resolve potential conflicts of opinion. Active conflict resolution practices can help identify potential biases and create a more inclusive dataset. 

\subsection{Data Science Considerations Toward Improving Data Inclusion and Quality}
In our survey study, participants indicated their applicable factors that should be considered to improve representation and inclusion in NLP datasets \textbf{[SQ-9a]}: annotation guidelines (63.83\%), identity inclusion datasets (61.70\%), data collection sources (57.45\%), model evaluation metrics (48.94\%), and privacy review (42.5\%). 

In addition, respondents suggested a few additional consideration factors, such as the need to expose models to diverse cultures and backgrounds, the incorporation of non-algorithmic metrics to assess bias and harm, questioning dataset ownership and access rights, and advocating for alternative annotation methods to enhance accuracy.

Focus group discussions delved deeper into the reasons and challenges behind the relatively lower emphasis on model evaluation metrics {\textbf{[FG-9a]}. Participant R1 noted, “A sentiment analysis algorithm might see a word like ``hate'' to be negative, but in the context of a statement like ‘I hate how this show ended on a cliffhanger' it is not negative". Furthermore, the lack of industry-wide standards in defining data quality and the success of NLP systems exacerbate this challenge (DA1, DA2). DE1 also emphasized the difficulty of designing metrics for voice-activated systems, which must account for accent and dialect diversity. 

The survey also explored the perceived responsibility of data science teams in developing systems that accurately represent users' experiences {\textbf{[SQ-9b]}. Respondents rated the importance on a 5-point Likert scale (0-5). Less than 5\% of NLP practitioners rated it slightly important, while a substantial majority (76.6\%) rated it either important or very important. 


\subsection{Opportunities for Public Participation in the NLP Dataset Production}
Our findings offer insight into the role in which the public can play when involved within various stages of the NLP data production pipeline. By examining survey responses and focus group discussions, we specifically focus on the implications for broadened public participation in enhancing the integrity of NLP datasets and system outcomes.

\subsubsection{Community Involvement in Data Production} Out of 47 respondents, 28 respondents (60\%) considered involving diverse community members with varying demographic backgrounds (gender, age, race/ethnicity, geo-location, etc.) in the NLP dataset production pipeline \textbf{[SQ-10a]}, while 19 respondents did not take it into account. Out of those 28 respondents, 18 detailed specific stages and methods to involve diverse communities.
Key areas identified include: 1) Early stage involvement (e.g., planning, project/study design), 2) Full-scale inclusion through the entire data production pipeline, 3) Specific stages for community input (e.g., collection, model training and evaluation, deployment and feedback, and 4) innovative equity-centered practices (e.g., community advisory board). Several respondents also highlighted challenges or the absence of diverse group involvement, often citing practical constraints related to their roles or existing project structures 
(e.g., “Not so far in my work, since data is usually already collected."). 

\subsubsection{Benefits of Diverse Community Involvement}
Focus group participants highlighted benefits of involving diverse external communities in NLP work \textbf{[FG-10a]}. DE1 noted that more diverse internal and external perspectives are often better equipped to address unique challenges, saying, ``To have the most well rounded product you need a well rounded team,'' while R1 emphasized co-annotation with end-users and other stakeholders, where community members collaborate with NLP practitioners to label, correct, and validate data, thereby enhancing data quality, increasing representation, and building users trust \& transparency. Participants noted the learning opportunities for NLP practitioners through community participatory innovation, including gaining deeper insights into different language use across communities; 
heightening awareness of ethical considerations (e.g., bias mitigation, data privacy); and learning effective ways to incorporate diverse feedback to ensure project success and relevance. By directly engaging diverse communities, as indicated in [SQ-7b] in NLP dataset production, practitioners and technology development organizations could ensure that the data collected is as representative and unbiased as possible. 
In fact, 96\% of survey respondents thought that the demographic background (gender, age, race, nationality, etc.) of data practitioners can affect the outcome of NLP datasets \textbf{[SQ-10b]}. This consensus underscores the pivotal role that diverse perspectives play in shaping data quality and model performance.

\subsubsection{Challenges in Engaging Community Data Agents} 
The survey evaluated the feasibility of integrating external communities into their NLP data annotation workflows on a 5-point Likert scale (0-5) \textbf{[SQ-10c]}. 16.02\% of NLP practitioners rated less feasible or not feasible, which was potentially caused by practical challenges or perceived structural, regulatory, or technical barriers. However, a significant majority ($83.98\%$) indicates some degree of feasibility (e.g., fairly feasible, feasible, very feasible) to involve external communities in the annotation process.

Respondents identified several potential challenges faced by external community members if involved in NLP data annotation \textbf{[SQ-10d]}: Access to data sources (44.68\%); Length of time to productively annotate (44.68\%); Understanding metrics of success (42.55\%);  Understanding the application or technology in which data is used for (40.43\%); Annotation quality (23.40\%). 7 respondents (14.89\%) indicated other challenges, including recruitment difficulties, logistical considerations, and integration with existing teams. Additional insights from open responses further underscored the complexity of these challenges and the practicalities of engaging external annotators or collaborators, such as model over-tuning, cultural and linguistic barriers, privacy and security concerns, and undervaluing external community experts knowledge.
The focus group elaborated on some underlying perceptions and structural issues in NLP data annotation that can impact the involvement and valuation of external data agents' skills and knowledge. Participant D1 stated “annotation and data work is considered and sold as low skill work gig/contract work that anyone could do.” 


 
\subsection{Perspectives on AI Policy and Regulation}
To gain perspective on the Blueprint for an AI Bill of Rights \cite{U.S.WhiteHouse_AIBillofRights_2022}, 
which is aimed at making automated systems work for the American people, we gauged practitioners’ familiarity with the Algorithmic Discrimination Protections (ADP) \textbf{ [SQ-11a]}. We found that 59.57\% of surveyed NLP practitioners reported low familiarity (e.g., slightly familiar, not familiar at all) and only a combined 14.89\% being very or extremely familiar, indicating a substantial knowledge gap regarding these protections and a need for broader educational initiatives within the industry.

Despite this gap, a substantial (93.62\%) of respondents rated the necessity of an AI Bill of Rights as at least moderately important, with a majority (62.71\%) leaning towards important or very important for preventing algorithmic discrimination.

\subsubsection{Importance of the AI Bill of Rights and Regulation}

Focus group participants underscored the importance of an AI Bill of Rights from both a technologist's perspective and the broader public's viewpoint. DE1 points out the practical and immediate consequences of AI advancements (e.g., creations of deepfakes, privacy invasions). They emphasize the need for regulation to keep people safe and mitigate harms already evident as AI becomes more pervasive and capable. 
DA2 advocates for the Bill as crucial for protecting consumer privacy and ensuring that data usage respects individual rights ``regardless of their class of life.'' Furthermore, DA2's call for equitable protections under AI regulations reinforces the role of AI policy and governance in promoting fairness and preventing discrimination.

\subsubsection{Necessary Algorithmic Discrimination Protections (ADP)} 27 survey respondents provided additional insights on the AI Bill of Rights Algorithmic Discrimination Protections \textbf{[SQ-11e]}. They highlighted their practical applicability, importance, and areas for improvement in NLP data-centric systems: 
\textit{1) Significance and Support for ADP:} Respondents' sentiments recognize ADP as fundamental in maintaining fairness and ethics in NLP practices (e.g., "ADP are very important...ensure that NLP systems are used in a way that is fair and equitable for all users"). Some responses emphasize the evolving nature of ADP and suggest that while current measures are a good start, ongoing refinement is necessary. 
\textit{2) Practical Applicability and Implementation Concerns:} Multiple respondents mention that implementing ADP policies is both crucial and challenging, advocating for robust, actionable strategies such as ``continuous auditing, transparency in model decisions, and proactive fairness measures in data collection'' to preemptively mitigate issues.
\textit{3) Skepticism and Concerns of Exploitative Practices:} Respondents highlighted a) trust issues on the effectiveness of government-led initiatives and b) inconsistent international protection standards.

Additionally, there were concerns that regulations might potentially drive exploitative practices or hinder open-source innovation: ``it's just going to force the people who don't agree with these rights to go underground, which leads to black market involvement.'' These insights indicate that while there is broad support for the ADP, there is a clear need for more concrete guidelines and effective implementation methods in the broader global AI landscape.

These findings underscore that while NLP practitioners are often expected to uphold fairness and equity standards, they operate within environments that frequently lack the organizational, technical, and governance support necessary to do so. The challenges they articulated—ranging from dataset limitations to institutional constraints—highlight a gap between ethical aspirations and practical realities. These tensions invite deeper examination of how responsibility is distributed across the AI development pipeline and raise important questions about the structural conditions required to meaningfully advance equitable NLP systems. We explore these implications further in the following discussion.

\section{Discussion}
Our study examines how NLP practitioners advance NLP data equity by reflecting on practices that produce representative, contextually accurate, and justice-oriented datasets to address technical mis-alignments and social inequities among diverse users.
This study offers three key contributions. First, it is among the few to center the voice of NLP practitioners -- capturing the lived realities of those embedded in the data production pipeline and their perceptions of equity, bias, and governance. Second, it connects these practitioner experiences directly to a multi-scalar governance framing, linking day-to-day decisions to institutional, national, and international policy contexts. Third, it advances participatory recommendations that bridge academic, industry, and civic domains, offering actionable pathways for enhancing practitioner agency and accountability.

\subsection{Expanding Metrics and Accountable Approaches to Fostering Diversity in Data Collection}

Developing robust strategies and metrics for measuring dataset diversity is essential. Further research emphasis could be placed on techniques for mitigating bias introduced during data collection. For instance, \citet{cunningham2024understanding} has proposed creating human-centered evaluation methods for better assessing AI’s sociocultural consequences in language technologies on underrepresented language communities. How are NLP professionals currently measuring the diversity of their datasets? Are there limitations to these metrics? Can new metrics be developed to provide a more comprehensive picture of dataset diversity? Simulations and user experience case studies can be used to evaluate the effectiveness of different bias mitigation strategies in various NLP data collection scenarios.

Subsequently, approaches to addressing the lack of diverse data sources (RQ1) can be extended to practices that enable transparency and equity in developing more representative NLP systems. Considering how might user-controlled mechanisms to collecting data improve the user experiences and performance of NLP systems. For instance, community-based governance and participatory data works can be integrated to broaden perspectives integral for humanistic data, especially from people of disproportionately affected communities. Community-based data governance and sovereignty empowers communities to actively participate in the data collection process and ensure that their experiences and perspectives are accurately represented \cite{mcnealy2020ecological, AdaLovelaceInstitute2021b, hovmand2014group}. Furthermore this enables specific communities-of-interest to reclaim authority and control over their data, ensuring self-determination in how their information is gathered, owned, stored, and utilized. This concept has proven to be especially crucial for First Nations or Indigenous peoples
\cite{lilley_maori_2024, west_data_2020}. Technical and policy frameworks for data co-operatives or data trusts \cite{mcnealy2020ecological, reed2019data} offer a way for stewards of community data to collectively own and govern their data, ensuring that they receive fair equity for their contributions and that their data are used in ways that benefit them \cite{pentland2021building, mcnealy2020ecological}.

\subsection{Promoting Community-collaborative Computing} 
While participation is recognized as essential for fairness and accountability in NLP, practical mechanisms for community participation remain limited. One promising model is the use of Community Advisory Boards (CABs), which have supported equitable algorithm development with underrepresented groups \cite{farmer_use_2022, mitchell_building_nodate}. In our study, one respondent proposed a CAB to incorporate Indigenous conceptions of health in clinical NLP. This also echoes recent participatory efforts to mitigate racial disparities in speech technologies \cite{cunningham_collaboratively_2023}.

Yet, in industry settings, such collaboration often depends on project leaders or even organizations who may de-prioritize public engagement due to perceived effort, costs, or misconceptions about skills necessary for data work. Investing in infrastructure to upskill community-based experts is often not considered. Although companies like Google, Meta, and Microsoft have launched expert-based external community-facing advisory boards \cite{smith2023equitable}, these mechanisms rarely involve experts whose language communities have been disproportionately impacted by NLP tools among other ML technologies.

In contrast, academic institutions face fewer barriers to sustained community collaboration. Initiatives like the Black Community Research Council at the University of Washington demonstrate how institutional research can meaningfully partner with marginalized communities to shape technology equity and design \cite{cunningham2024methods}.

\subsection{Equipping Practitioners for Equity-Centered NLP Work}

Despite interest in fairness, many NLP practitioners lack experience in core equity practices. Only 23.4\% of survey respondents reported any prior engagement with dataset bias mitigation, revealing substantial gaps in both training and institutional support. This gap reflects both workplace dynamics and the absence of critical education in mainstream technical training.

To address this, academic and professional training programs must more explicitly integrate social and ethical dimensions of computing. For example, ``CompSci 240: Race, Gender, Class, \& Computing'' at Duke University trains students to interrogate how identity, power, and systems of inequality shape the development and deployment of AI \cite{CompSci240}. Industry initiatives, such as Google’s Product Inclusion Summit, represent a promising model for skills-building in inclusive design.

However, training alone is not sufficient. Many participants expressed skepticism toward checklist-style diversity efforts that emphasize representational inclusion without substantive engagement. These concerns align with critiques of ``diversity-washing'' \cite{whitney2024diversitywashing, dalal2024provocation}, which caution that superficial inclusion efforts can obscure structural inequities or bypass community consent. Practitioners emphasized that equity must extend beyond demographic representation to include epistemic justice, participatory design, and reciprocal engagement with affected communities. Taken together, these insights underscore the importance of coupling practitioner capacity-building with structural reforms. Equity in NLP is not only a technical or moral challenge, but an institutional one—requiring both governance support and practitioner empowerment.
\subsection{Resituating Practitioner Agency in Multi-Scalar Governance}

Our findings reveal that while NLP practitioners play a crucial role in shaping equitable AI outcomes, their efforts are often constrained by organizational priorities, short-term incentives, and a lack of structural support or accountability mechanisms. These tensions highlight the need to move beyond framing NLP data equity as an individual responsibility and instead embed it within broader AI governance ecosystems.

Drawing on Drahos’ Regulatory Theory \cite{drahos2017regulatory}, we argue for multi-scalar approaches to AI governance—those that span policy regimes, institutional norms, and technical infrastructures. Survey responses, including from items [\textbf{SQ-11a}] and [\textbf{SQ-11d}], showed limited familiarity with algorithmic discrimination protections but a strong consensus that public policymakers, rather than private companies, should lead in ensuring fairness and safety \cite{lander2021americans, araya2020renewing}. Calls for international alignment further resonate with UNESCO's recommendation for globally harmonized AI ethics.

Public-private collaboration has a role to play, but only if accountability and public interest are centered \cite{crawford2021atlas}. Civic organizations (e.g., ACLU) can help bridge this gap by facilitating community engagement beyond corporate priorities \cite{costanza-chock_design_2018, brown_emergent_2017}. Governance frameworks must explicitly support practitioner agency—not simply by prescribing compliance, but by providing the institutional scaffolding needed to make equity work viable and meaningful within teams and workflows.

\section{Study Limitations} 
While this study employed a mixed‑methods approach to provide both breadth and depth of insight, several limitations should be noted. First, the survey and focus groups were conducted in late 2023 and early 2024, ensuring relevance to shifts in NLP data practices following the proliferation of large language models (LLMs). Second, our participant sample over-represents Black practitioners, reflecting targeted outreach within AI fairness and equity networks. While this offers critical insights from often underrepresented voices, the findings are qualitative and exploratory rather than statistically generalizable. We acknowledge the potential for response bias, as participants may have been predisposed to engage with NLP data equity issues. Third, while our study focused on practitioners' individual perspectives, we recognize that organizational structures and workplace policies play a significant role in shaping NLP data equity practices. A deeper analysis of these dynamics was beyond the study’s scope expanded in the discussion with reference to prior research. Finally, the study is primarily U.S. centric and findings should not be assumed to reflect global NLP practitioner contexts.

\section{Conclusion}
This study examined how NLP practitioners navigate NLP data equity in the design and development of language technologies. Through a mixed-methods study of U.S.-based practitioners, we surfaced how individuals responsible for dataset development and annotation engage with ethical frameworks, fairness goals, and systemic constraints. Our findings reveal persistent tensions between commercial imperatives and equity commitments, as well as practitioner skepticism toward superficial inclusion efforts.

We argue that equitable NLP systems cannot be achieved solely through technical interventions or individual initiative. Instead, they require structural realignments—governance frameworks that support practitioner agency, participatory infrastructure for community engagement, and accountability mechanisms embedded across the AI development lifecycle. Drawing on critiques of diversity-washing and multi-scalar governance theory, we advocate for a shift from representational metrics to relational approaches grounded in consent, context, and community-defined equity. By foregrounding practitioner perspectives, this work contributes to ongoing efforts to rethink responsibility and accountability in AI development—not as abstract principles, but as situated practices shaped by institutional conditions and socio-technical dynamics.

\section*{Acknowledgements}
This work was supported by the National Science Foundation Graduate Research Fellowship, the Google Award for Inclusive Research, and the Social Science Research Council Just Tech Fellowship. The first author gratefully acknowledges the commitment of these organizations to advancing research at the intersection of technology equity and social justice.

\bibliography{aaai25}

\begin{thebibliography}{71}
\providecommand{\natexlab}[1]{#1}

\bibitem[{U.S(2022)}]{U.S.WhiteHouse_AIBillofRights_2022}
 2022.
\newblock Blueprint for an AI Bill of Rights (OSTP).

\bibitem[{Act(2024)}]{act2024eu}
Act, E. A.~I. 2024.
\newblock The eu artificial intelligence act.
\newblock \emph{European Union}.

\bibitem[{{Ada Lovelace Institute}(2021)}]{AdaLovelaceInstitute2021b}
{Ada Lovelace Institute}. 2021.
\newblock Participatory Data Stewardship.

\bibitem[{Araya and Nieto-G{\'o}mez(2020)}]{araya2020renewing}
Araya, D.; and Nieto-G{\'o}mez, R. 2020.
\newblock Renewing multilateral governance in the age of AI.
\newblock \emph{MODERN CONFLICT AND ARTIFICIAL INTELLIGENCE}, 6.

\bibitem[{Barocas, Hardt, and Narayanan(2020)}]{barocas2020fairness}
Barocas, S.; Hardt, M.; and Narayanan, A. 2020.
\newblock Fairness and machine learning.
\newblock \emph{Recommender systems handbook}, 1: 453--459.

\bibitem[{Bender and Friedman(2018)}]{bender-friedman-2018-data}
Bender, E.~M.; and Friedman, B. 2018.
\newblock Data Statements for Natural Language Processing: Toward Mitigating System Bias and Enabling Better Science.
\newblock \emph{Transactions of the Association for Computational Linguistics}, 6: 587--604.

\bibitem[{Braun and Clarke(2006-01)}]{braun_using_2006}
Braun, V.; and Clarke, V. 2006-01.
\newblock Using thematic analysis in psychology.
\newblock 3(2): 77--101.

\bibitem[{brown(2017)}]{brown_emergent_2017}
brown, a.~m. 2017.
\newblock \emph{Emergent Strategy: Shaping Change, Changing Worlds}.
\newblock {AK} Press.
\newblock ISBN 978-1-84935-261-1.

\bibitem[{Costanza-Chock(2018)}]{costanza-chock_design_2018}
Costanza-Chock, S. 2018.
\newblock Design Justice: Towards an Intersectional Feminist Framework for Design Theory and Practice.
\newblock ({ID} 3189696).

\bibitem[{Crawford(2021)}]{crawford2021atlas}
Crawford, K. 2021.
\newblock The Atlas of AI: Power, Politics, and the Planetary Costs of Artificial Intelligence.

\bibitem[{Cunningham et~al.(2024)Cunningham, Blodgett, Madaio, Iii, Harrington, and Wallach}]{cunningham2024understanding}
Cunningham, J.; Blodgett, S.~L.; Madaio, M.; Iii, H.~D.; Harrington, C.; and Wallach, H. 2024.
\newblock Understanding the impacts of language technologies’ performance disparities on African American language speakers.
\newblock In \emph{Findings of the Association for Computational Linguistics ACL 2024}, 12826--12833.

\bibitem[{Cunningham(2023)}]{cunningham_collaboratively_2023}
Cunningham, J.~L. 2023.
\newblock Collaboratively Mitigating Racial Disparities in Automated Speech Recognition and Language Technologies with African American English Speakers: Community-Collaborative and Equity-Centered Approaches Toward Designing Inclusive Natural Language Systems.
\newblock In \emph{Extended Abstracts of the 2023 {CHI} Conference on Human Factors in Computing Systems}, {CHI} {EA} '23, 1--5. Association for Computing Machinery.
\newblock ISBN 978-1-4503-9422-2.

\bibitem[{Cunningham(2024)}]{cunningham2024methods}
Cunningham, J.~L. 2024.
\newblock \emph{Methods of Designing Justice-oriented Interactive AI Systems}.
\newblock Ph.D. thesis, University of Washington.

\bibitem[{Dalal, Hall, and Johnson(2024)}]{dalal2024provocation}
Dalal, S.; Hall, S.~M.; and Johnson, N. 2024.
\newblock Provocation: Who benefits from" inclusion" in Generative AI?
\newblock \emph{arXiv preprint arXiv:2411.09102}.

\bibitem[{Deas et~al.(2023)Deas, Grieser, Kleiner, Patton, Turcan, and {McKeown}}]{deas_evaluation_2023}
Deas, N.; Grieser, J.; Kleiner, S.; Patton, D.; Turcan, E.; and {McKeown}, K. 2023.
\newblock Evaluation of African American Language Bias in Natural Language Generation.
\newblock 2305.14291 [cs].

\bibitem[{Deng et~al.(2023)Deng, Yildirim, Chang, Eslami, Holstein, and Madaio}]{10.1145/3593013.3594037}
Deng, W.~H.; Yildirim, N.; Chang, M.; Eslami, M.; Holstein, K.; and Madaio, M. 2023.
\newblock Investigating Practices and Opportunities for Cross-functional Collaboration around AI Fairness in Industry Practice.
\newblock In \emph{Proceedings of the 2023 ACM Conference on Fairness, Accountability, and Transparency}, FAccT '23, 705–716. New York, NY, USA: Association for Computing Machinery.
\newblock ISBN 9798400701924.

\bibitem[{Deng et~al.(2024)Deng, Qiu, Tan, Pan, Jue, Fang, Xu, Chu, and Qi}]{deng2024promoting}
Deng, Y.; Qiu, X.; Tan, X.; Pan, J.; Jue, C.; Fang, Z.; Xu, Y.; Chu, W.; and Qi, Y. 2024.
\newblock Promoting Equality in Large Language Models: Identifying and Mitigating the Implicit Bias based on Bayesian Theory.
\newblock \emph{arXiv preprint arXiv:2408.10608}.

\bibitem[{Denton et~al.(2021)Denton, Díaz, Kivlichan, Prabhakaran, and Rosen}]{denton_whose_2021}
Denton, R.; Díaz, M.; Kivlichan, I.; Prabhakaran, V.; and Rosen, R. 2021.
\newblock Whose Ground Truth? Accounting for Individual and Collective Identities Underlying Dataset Annotation.
\newblock 2112.04554 [cs].

\bibitem[{Dourish(2017)}]{dourish2017stuff}
Dourish, P. 2017.
\newblock \emph{The Stuff of Bits: An Essay on the Materialities of Information}.
\newblock MIT Press.

\bibitem[{Drahos(2017)}]{drahos2017regulatory}
Drahos, P. 2017.
\newblock \emph{Regulatory theory: Foundations and applications}.
\newblock ANU Press.

\bibitem[{Farmer et~al.(2022)Farmer, Osei~Baah, Williams, Ortiz-Chapparo, Mitchell, Jackson, Collins, Graham, Wallen, Powell-Wiley, and Johnson}]{farmer_use_2022}
Farmer, N.; Osei~Baah, F.; Williams, F.; Ortiz-Chapparo, E.; Mitchell, V.~M.; Jackson, L.; Collins, B.; Graham, L.; Wallen, G.~R.; Powell-Wiley, T.~M.; and Johnson, A. 2022.
\newblock Use of a community advisory board to build equitable algorithms for participation in clinical trials: a protocol paper for {HoPeNET}.
\newblock 29(1): e100453.

\bibitem[{Floridi et~al.(2018)Floridi, Cowls, Beltrametti, Chatila, Chazerand, Dignum, Luetge, Madelin, Pagallo, Rossi et~al.}]{floridi2018ai4people}
Floridi, L.; Cowls, J.; Beltrametti, M.; Chatila, R.; Chazerand, P.; Dignum, V.; Luetge, C.; Madelin, R.; Pagallo, U.; Rossi, F.; et~al. 2018.
\newblock AI4People—an ethical framework for a good AI society: opportunities, risks, principles, and recommendations.
\newblock \emph{Minds and machines}, 28: 689--707.

\bibitem[{Gebru et~al.(2021-12)Gebru, Morgenstern, Vecchione, Vaughan, Wallach, Iii, and Crawford}]{gebru_datasheets_2021}
Gebru, T.; Morgenstern, J.; Vecchione, B.; Vaughan, J.~W.; Wallach, H.; Iii, H.~D.; and Crawford, K. 2021-12.
\newblock Datasheets for datasets.
\newblock 64(12): 86--92.

\bibitem[{Goyal et~al.(2022)Goyal, Kivlichan, Rosen, and Vasserman}]{goyal_is_2022}
Goyal, N.; Kivlichan, I.; Rosen, R.; and Vasserman, L. 2022.
\newblock Is Your Toxicity My Toxicity? Exploring the Impact of Rater Identity on Toxicity Annotation.
\newblock 2205.00501 [cs].

\bibitem[{Hanna et~al.(2020)Hanna, Denton, Smart, and Smith-Loud}]{hanna_towards_2020}
Hanna, A.; Denton, E.; Smart, A.; and Smith-Loud, J. 2020.
\newblock Towards a critical race methodology in algorithmic fairness.
\newblock In \emph{Proceedings of the 2020 Conference on Fairness, Accountability, and Transparency}, {FAT}* '20, 501--512. Association for Computing Machinery.
\newblock ISBN 978-1-4503-6936-7.

\bibitem[{Hill et~al.(2022)Hill, Ogle, Gottlieb, Santen, and Artino~Jr}]{hill_educators_2022}
Hill, J.; Ogle, K.; Gottlieb, M.; Santen, S.~A.; and Artino~Jr, A.~R. 2022.
\newblock Educator's blueprint: A how-to guide for collecting validity evidence in survey- based research.
\newblock 6(6): e10835.
\newblock \_eprint: https://onlinelibrary.wiley.com/doi/pdf/10.1002/aet2.10835.

\bibitem[{Holstein et~al.(2019)Holstein, Wortman~Vaughan, Daum{\'e}~III, Dudik, and Wallach}]{holstein2019improving}
Holstein, K.; Wortman~Vaughan, J.; Daum{\'e}~III, H.; Dudik, M.; and Wallach, H. 2019.
\newblock Improving fairness in machine learning systems: What do industry practitioners need?
\newblock In \emph{Proceedings of the 2019 CHI conference on human factors in computing systems}, 1--16.

\bibitem[{Hovmand and Hovmand(2014)}]{hovmand2014group}
Hovmand, P.~S.; and Hovmand, P.~S. 2014.
\newblock \emph{Group model building and community-based system dynamics process}.
\newblock Springer.

\bibitem[{Hutchinson et~al.(2020)Hutchinson, Prabhakaran, Denton, Webster, Zhong, and Denuyl}]{hutchinson-etal-2020-social}
Hutchinson, B.; Prabhakaran, V.; Denton, E.; Webster, K.; Zhong, Y.; and Denuyl, S. 2020.
\newblock Social Biases in {NLP} Models as Barriers for Persons with Disabilities.
\newblock In Jurafsky, D.; Chai, J.; Schluter, N.; and Tetreault, J., eds., \emph{Proceedings of the 58th Annual Meeting of the Association for Computational Linguistics}, 5491--5501. Online: Association for Computational Linguistics.

\bibitem[{Irani(2016)}]{irani2016hackathons}
Irani, L. 2016.
\newblock Hackathons and the making of entrepreneurial citizenship.
\newblock \emph{Science, Technology, \& Human Values}, 41(5): 705--740.

\bibitem[{Jo and Gebru(2020)}]{jo_lessons_2020}
Jo, E.~S.; and Gebru, T. 2020.
\newblock Lessons from archives: strategies for collecting sociocultural data in machine learning.
\newblock In \emph{Proceedings of the 2020 Conference on Fairness, Accountability, and Transparency}, {FAT}* '20, 306--316. Association for Computing Machinery.
\newblock ISBN 978-1-4503-6936-7.

\bibitem[{Jobin, Ienca, and Vayena(2019)}]{jobin2019global}
Jobin, A.; Ienca, M.; and Vayena, E. 2019.
\newblock The global landscape of AI ethics guidelines.
\newblock \emph{Nature machine intelligence}, 1(9): 389--399.

\bibitem[{Joshi et~al.(2020)Joshi, Santy, Budhiraja, Bali, and Choudhury}]{joshi2020state}
Joshi, P.; Santy, S.; Budhiraja, A.; Bali, K.; and Choudhury, M. 2020.
\newblock The state and fate of linguistic diversity and inclusion in the NLP world.
\newblock \emph{arXiv preprint arXiv:2004.09095}.

\bibitem[{Kamikubo et~al.(2022)Kamikubo, Wang, Marte, Mahmood, and Kacorri}]{kamikubo_data_2022}
Kamikubo, R.; Wang, L.; Marte, C.; Mahmood, A.; and Kacorri, H. 2022.
\newblock Data Representativeness in Accessibility Datasets: A Meta-Analysis.
\newblock In \emph{Proceedings of the 24th International {ACM} {SIGACCESS} Conference on Computers and Accessibility}, {ASSETS} '22, 1--15. Association for Computing Machinery.
\newblock ISBN 978-1-4503-9258-7.

\bibitem[{Kapania, Taylor, and Wang(2023)}]{kapania_hunt_2023}
Kapania, S.; Taylor, A.~S.; and Wang, D. 2023.
\newblock A hunt for the Snark: Annotator Diversity in Data Practices.
\newblock In \emph{Proceedings of the 2023 {CHI} Conference on Human Factors in Computing Systems}, {CHI} '23, 1--15. Association for Computing Machinery.
\newblock ISBN 978-1-4503-9421-5.

\bibitem[{Koenecke et~al.(2020)Koenecke, Nam, Lake, Nudell, Quartey, Mengesha, Toups, Rickford, Jurafsky, and Goel}]{koenecke_racial_2020}
Koenecke, A.; Nam, A.; Lake, E.; Nudell, J.; Quartey, M.; Mengesha, Z.; Toups, C.; Rickford, J.~R.; Jurafsky, D.; and Goel, S. 2020.
\newblock Racial disparities in automated speech recognition.
\newblock 117(14): 7684--7689.

\bibitem[{Lander and Nelson(2021)}]{lander2021americans}
Lander, E.; and Nelson, A. 2021.
\newblock Americans need a Bill of rights for an AI-powered world. Wired.

\bibitem[{Lazar and Nelson(2023)}]{lazar2023ai}
Lazar, S.; and Nelson, A. 2023.
\newblock AI safety on whose terms?

\bibitem[{Lilley et~al.(2024)Lilley, Oliver, Cranefield, and Lewellen}]{lilley_maori_2024}
Lilley, S.; Oliver, G.; Cranefield, J.; and Lewellen, M. 2024.
\newblock Māori data sovereignty: contributions to data cultures in the government sector in New Zealand.
\newblock 0(0): 1--16.
\newblock Publisher: Routledge \_eprint: https://doi.org/10.1080/1369118X.2024.2302987.

\bibitem[{Lincoln(1985)}]{lincoln1985naturalistic}
Lincoln, Y.~S. 1985.
\newblock \emph{Naturalistic inquiry}, volume~75.
\newblock sage.

\bibitem[{Marshall and Jonker(2011)}]{marshall2011introduction}
Marshall, G.; and Jonker, L. 2011.
\newblock An introduction to inferential statistics: A review and practical guide.
\newblock \emph{Radiography}, 17(1): e1--e6.

\bibitem[{Mathur et~al.(2022)Mathur, Rentsch, Venkataraman, Fatumo, Jobe, Angkurawaranon, Ong, Wong, and Siddiqui}]{mathur2022we}
Mathur, R.; Rentsch, C.~T.; Venkataraman, K.; Fatumo, S.; Jobe, M.; Angkurawaranon, C.; Ong, S.~E.; Wong, A.~Y.; and Siddiqui, M.~K. 2022.
\newblock How do we collect good-quality data on race and ethnicity and address the trust gap?
\newblock \emph{The Lancet}, 400(10368): 2028--2030.

\bibitem[{McConvey and Guha(2024)}]{mcconvey2024not}
McConvey, K.; and Guha, S. 2024.
\newblock " This is not a data problem": Algorithms and Power in Public Higher Education in Canada.
\newblock In \emph{Proceedings of the CHI Conference on Human Factors in Computing Systems}, 1--14.

\bibitem[{McNealy(2020)}]{mcnealy2020ecological}
McNealy, J. 2020.
\newblock An Ecological Approach to Data Governance.
\newblock \emph{Bulletin of the American Physical Society}, 65.

\bibitem[{Miceli, Schuessler, and Yang(2020)}]{miceli_between_2020}
Miceli, M.; Schuessler, M.; and Yang, T. 2020.
\newblock Between Subjectivity and Imposition: Power Dynamics in Data Annotation for Computer Vision.
\newblock 4: 115:1--115:25.

\bibitem[{Miceli et~al.(2022)Miceli, Yang, Alvarado~Garcia, Posada, Wang, Pohl, and Hanna}]{miceli_documenting_2022}
Miceli, M.; Yang, T.; Alvarado~Garcia, A.; Posada, J.; Wang, S.~M.; Pohl, M.; and Hanna, A. 2022.
\newblock Documenting Data Production Processes: A Participatory Approach for Data Work.
\newblock 6: 510:1--510:34.

\bibitem[{Mitchell et~al.()Mitchell, Perry, Rorai, Ilardo, Lichtenberg, and Jackson}]{mitchell_building_nodate}
Mitchell, J.; Perry, T.; Rorai, V.; Ilardo, J.; Lichtenberg, P.~A.; and Jackson, J.~S. ????
\newblock Building and Sustaining a Community Advisory Board of African American Older Adults as the Foundation for Volunteer Research Recruitment and Retention in Health Sciences.
\newblock 30: 755--764.

\bibitem[{Mothilal, Guha, and Ahmed(2024)}]{mothilal2024towards}
Mothilal, R.~K.; Guha, S.; and Ahmed, S.~I. 2024.
\newblock Towards a Non-Ideal Methodological Framework for Responsible ML.
\newblock \emph{arXiv preprint arXiv:2401.11131}.

\bibitem[{Muller et~al.(2019)Muller, Lange, Wang, Piorkowski, Tsay, Liao, Dugan, and Erickson}]{muller2019data}
Muller, M.; Lange, I.; Wang, D.; Piorkowski, D.; Tsay, J.; Liao, Q.~V.; Dugan, C.; and Erickson, T. 2019.
\newblock How data science workers work with data: Discovery, capture, curation, design, creation.
\newblock In \emph{Proceedings of the 2019 CHI conference on human factors in computing systems}, 1--15.

\bibitem[{Muller and Kuhn(1993)}]{muller1993participatory}
Muller, M.~J.; and Kuhn, S. 1993.
\newblock Participatory design.
\newblock \emph{Communications of the ACM}, 36(6): 24--28.

\bibitem[{Olson and Kellogg(2014)}]{olson2014ways}
Olson, J.~S.; and Kellogg, W.~A. 2014.
\newblock \emph{Ways of Knowing in HCI}, volume~2.
\newblock Springer.

\bibitem[{Pang et~al.(2023)Pang, Cenatempo, Graham, Kuehn, Whisenant, Botchway, Stone~Perez, and Koenecke}]{pang_auditing_2023}
Pang, R.~Y.; Cenatempo, J.; Graham, F.; Kuehn, B.; Whisenant, M.; Botchway, P.; Stone~Perez, K.; and Koenecke, A. 2023.
\newblock Auditing Cross-Cultural Consistency of Human-Annotated Labels for Recommendation Systems.
\newblock In \emph{Proceedings of the 2023 {ACM} Conference on Fairness, Accountability, and Transparency}, {FAccT} '23, 1531--1552. Association for Computing Machinery.
\newblock ISBN 9798400701924.

\bibitem[{Paullada et~al.(2021)Paullada, Raji, Bender, Denton, and Hanna}]{paullada2021data}
Paullada, A.; Raji, I.~D.; Bender, E.~M.; Denton, E.; and Hanna, A. 2021.
\newblock Data and its (dis) contents: A survey of dataset development and use in machine learning research.
\newblock \emph{Patterns}, 2(11).

\bibitem[{Pentland, Lipton, and Hardjono(2021)}]{pentland2021building}
Pentland, A.; Lipton, A.; and Hardjono, T. 2021.
\newblock \emph{Building the new economy: Data as capital}.
\newblock MIT Press.

\bibitem[{Raji et~al.(2021)Raji, Bender, Paullada, Denton, and Hanna}]{raji_ai_2021}
Raji, I.~D.; Bender, E.~M.; Paullada, A.; Denton, E.; and Hanna, A. 2021.
\newblock {AI} and the Everything in the Whole Wide World Benchmark.
\newblock 2111.15366 [cs].

\bibitem[{Reed and Ng(2019)}]{reed2019data}
Reed, C.; and Ng, I. 2019.
\newblock Data trusts as an AI governance mechanism.
\newblock \emph{Available at SSRN 3334527}.

\bibitem[{Rothschild et~al.(2022)Rothschild, Meng, DiSalvo, Johnson, Shapiro, and DiSalvo}]{rothschild2022interrogating}
Rothschild, A.; Meng, A.; DiSalvo, C.; Johnson, B.; Shapiro, B.~R.; and DiSalvo, B. 2022.
\newblock Interrogating data work as a community of practice.
\newblock \emph{Proceedings of the ACM on Human-Computer Interaction}, 6(CSCW2): 1--28.

\bibitem[{Sambasivan et~al.(2021)Sambasivan, Arnesen, Hutchinson, Doshi, and Prabhakaran}]{sambasivan_re-imagining_2021}
Sambasivan, N.; Arnesen, E.; Hutchinson, B.; Doshi, T.; and Prabhakaran, V. 2021.
\newblock Re-imagining Algorithmic Fairness in India and Beyond.
\newblock In \emph{Proceedings of the 2021 {ACM} Conference on Fairness, Accountability, and Transparency}, {FAccT} '21, 315--328. Association for Computing Machinery.
\newblock ISBN 978-1-4503-8309-7.

\bibitem[{Sap et~al.(2019{\natexlab{a}})Sap, Card, Gabriel, Choi, and Smith}]{sap_risk_2019}
Sap, M.; Card, D.; Gabriel, S.; Choi, Y.; and Smith, N.~A. 2019{\natexlab{a}}.
\newblock The Risk of Racial Bias in Hate Speech Detection.
\newblock In \emph{Proceedings of the 57th Annual Meeting of the Association for Computational Linguistics}, 1668--1678. Association for Computational Linguistics.

\bibitem[{Sap et~al.(2019{\natexlab{b}})Sap, Card, Gabriel, Choi, and Smith}]{sap2019risk}
Sap, M.; Card, D.; Gabriel, S.; Choi, Y.; and Smith, N.~A. 2019{\natexlab{b}}.
\newblock The risk of racial bias in hate speech detection.
\newblock In \emph{Proceedings of the 57th annual meeting of the association for computational linguistics}, 1668--1678.

\bibitem[{Sloan et~al.(2020)Sloan, Moss, Awomolo, and Forlano}]{sloan2020participation}
Sloan, M.; Moss, E.; Awomolo, O.; and Forlano, L. 2020.
\newblock Participation is not a design fix for machine learning (pp. 1--7).
\newblock In \emph{Proceedings of the International Conference on Machine Learning, Vienna, Austria}.

\bibitem[{Sloane et~al.(2020)Sloane, Moss, Awomolo, and Forlano}]{sloane_participation_nodate}
Sloane, M.; Moss, E.; Awomolo, O.; and Forlano, L. 2020.
\newblock Participation is not a Design Fix for Machine Learning.
\newblock 7.

\bibitem[{Smith-Loud et~al.(2023)Smith-Loud, Smart, Neal, Ebinama, Corbett, Nicholas, Rashid, Peckham, Murphy-Gray, Morris et~al.}]{smith2023equitable}
Smith-Loud, J.; Smart, A.; Neal, D.; Ebinama, A.; Corbett, E.; Nicholas, P.; Rashid, Q.; Peckham, A.; Murphy-Gray, S.; Morris, N.; et~al. 2023.
\newblock The Equitable AI Research Roundtable (EARR): Towards Community-Based Decision Making in Responsible AI Development.
\newblock \emph{arXiv preprint arXiv:2303.08177}.

\bibitem[{Suchman(2002)}]{suchman_located_2002}
Suchman, L. 2002.
\newblock Located accountabilities in technology production.
\newblock 14: 16.

\bibitem[{Tatman(2017)}]{tatman_gender_2017}
Tatman, R. 2017.
\newblock Gender and Dialect Bias in {YouTube}'s Automatic Captions.
\newblock In \emph{Proceedings of the First {ACL} Workshop on Ethics in Natural Language Processing}, 53--59. Association for Computational Linguistics.

\bibitem[{Thakkar et~al.(2022)Thakkar, Ismail, Kumar, Hanna, Sambasivan, and Kumar}]{thakkar2022machine}
Thakkar, D.; Ismail, A.; Kumar, P.; Hanna, A.; Sambasivan, N.; and Kumar, N. 2022.
\newblock When is machine learning data good?: Valuing in public health datafication.
\newblock In \emph{Proceedings of the 2022 CHI Conference on Human Factors in Computing Systems}, 1--16.

\bibitem[{University(2021)}]{CompSci240}
University, D. 2021.
\newblock CompSci 240: Race, Gender, Class, \& Computing.
\newblock Course offered at Duke University.

\bibitem[{West, Hudson, and Kukutai(2020)}]{west_data_2020}
West, K.; Hudson, M.; and Kukutai, T. 2020.
\newblock Data Ethics and Data Governance from A Māori World View.
\newblock In George, L.; Tauri, J.; and Te~Ata~o Tu~{MacDonald}, L., eds., \emph{Indigenous Research Ethics: Claiming Research Sovereignty Beyond Deficit and the Colonial Legacy}, volume~6 of \emph{Advances in Research Ethics and Integrity}, 67--81. Emerald Publishing Limited.
\newblock ISBN 978-1-78769-390-6 978-1-78769-389-0.

\bibitem[{Whitney and Norman(2024)}]{whitney2024diversitywashing}
Whitney, C.~D.; and Norman, J. 2024.
\newblock Real Risks of Fake Data: Synthetic Data, Diversity-Washing and Consent Circumvention.
\newblock In \emph{Proceedings of the 2024 ACM Conference on Fairness, Accountability, and Transparency}, FAccT '24, 1733–1744. New York, NY, USA: Association for Computing Machinery.
\newblock ISBN 9798400704505.

\bibitem[{Widder and Nafus(2023)}]{widder2023dislocated}
Widder, D.~G.; and Nafus, D. 2023.
\newblock Dislocated accountabilities in the “AI supply chain”: Modularity and developers’ notions of responsibility.
\newblock \emph{Big Data \& Society}, 10(1): 20539517231177620.

\bibitem[{Ziman et~al.(2018)Ziman, Heusser, Fitzpatrick, Field, and Manning}]{ziman2018automatic}
Ziman, K.; Heusser, A.~C.; Fitzpatrick, P.~C.; Field, C.~E.; and Manning, J.~R. 2018.
\newblock Is automatic speech-to-text transcription ready for use in psychological experiments?
\newblock \emph{Behavior research methods}, 50: 2597--2605.

\end{thebibliography}

\appendix
\section{Appendices}
\subsection{A. Survey Questions}
\textbf{Pre-Screener Questionnaire.}
\begin{enumerate}
\item Do you have experience in roles/projects related to the production of text or speech datasets for Natural Language Processing (NLP) systems or technologies?\textsuperscript{*} \\
(Description: To be eligible for participation in this study, you must identify as a data practitioner (e.g. data scientists, data engineers, researchers, contract-data workers, or other data practitioners) engaged in the aforementioned fields or related areas.
\begin{enumerate}
\item[A.] Yes
\item[B.] No (Ends survey)
\item[C.] I'm not sure
\end{enumerate}
\item Please specify your relevant professional or preferred title and area of work/expertise. No mention of the workplace is required.\textsuperscript{*}
(e.g. Researcher, NLP Speech Recognition for Voice Assistants)
\item Is your employment/work based in the U.S?\textsuperscript{*}
(Description: To be eligible for participation in this study, you must be employed/working based in the U.S.)
\begin{enumerate}
\item[A.] Yes
\item[B.] No (Ends survey)
\end{enumerate}
\item Please indicate your age.\textsuperscript{*} (If the participant enters anything lower than 18, the survey will be submitted.)
(Description: As representative of today's date. To be eligible for participation in this study, you must be 18 years of age and older.)
\end{enumerate}
\textbf{Experience in Natural Language Processing (NLP).}
\begin{enumerate}
\item {[SQ-6a]} What experiences have you had within data science for text or speech NLP? (check-box)
\begin{enumerate}
\item[A.] data collection or annotation for text/speech model training
\item[B.] research on data practices or methods
\item[C.] organizing and managing data annotation projects
\item[D.] training and supervising teams of data workers
\item[E.] auditing or evaluating quality of data collection/annotation
\item[F.] mitigating bias in data training sets
\item[G.] developing evaluation models to measure effectiveness of training data
\item[H.] others
\end{enumerate}
\item {[SQ-6b]} What is your expertise level in data practices (organizing, collecting, transcribing, annotating, or analyzing) text or speech data for NLP datasets? *
\begin{enumerate}
\item[A.] Beginner: I have limited experience with these data practices for NLP projects
\item[B.] Intermediate: I have some experience and have worked on a few NLP projects involving data practices
\item[C.] Advanced: I have significant experience in working on complex NLP projects and have a deep understanding of data practices
\item[D.] Expert: I have extensive expertise in data practices for NLP projects and have led people in a wide range of tasks and project efforts
\item[E.] N/A: My role does not involve using data practices in producing text or speech data for NLP datasets
\end{enumerate}
\item {[SQ-6c]} Can you share an example of a specific text or speech NLP-enabled project (technology, feature, tool) that you have worked on? (e.g. virtual assistant, smart speaker, smart home devices, speech-to-text dictation apps)? (short answer)
\end{enumerate}

\textbf{Data Collection.}
\begin{enumerate}
\item {[SQ-7a]} How do you strive for diverse representation and inclusion in the datasets you curate?
\item {[SQ-7b]} Do you take into account any social factors (ethnic, cultural, or linguistic) of data subjects during NLP data collection? Please elaborate on how this is recognized and implemented.
\end{enumerate}

\textbf{Consensus Building and Quality Assurance in Data Annotation.}
\begin{enumerate}
\item {[SQ-8a]} In your project experience, how is a data annotator's performance measured?
\begin{enumerate}
\item[A.] Accuracy of Annotations
\item[B.] Speed of Annotation Completion
\item[C.] Quality of Annotation
\item[D.] Feedback from other Annotators
\item[E.] Compliance with Annotation Guidelines
\item[F.] Others
\item[G.] Not Applicable
\end{enumerate}
\item {[SQ-8b]} In your project experience(s), how much time is typically allocated to resolving potential conflicts of opinion among data annotators?
\begin{enumerate}
\item[A.] None: No time is allocated to resolving potential conflicts of opinion.
\item[B.] Very little: Only a small amount of time is dedicated to addressing conflicts of opinion.
\item[C.] Some: A moderate portion of time is allocated to resolving potential conflicts of opinion.
\item[D.] Significant: A substantial amount of time is dedicated to addressing conflicts of opinion.
\item[E.] N/A
\end{enumerate}
\end{enumerate}

\textbf{Considerations Toward Representation and Inclusion.}
\begin{enumerate}
\item {[SQ-9a]} What factors should be taken into account to improve representation and inclusion in NLP datasets? * (Multi-select)
\begin{enumerate}
\item[A.] Identity Representation Datasets: Ensure the annotated dataset includes a diverse range of user groups, cultural backgrounds, and linguistic variation to avoid biases and under-representation.
\item[B.] Annotation Guidelines: Develop comprehensive annotation guidelines that consider the sensitivities and nuances of different cultures and languages to avoid offensive or inappropriate content.
\item[C.] Data Collection Sources: Utilize multiple and varied sources for data collection, spanning across different regions and communities, to capture a more comprehensive understanding of language usage.
\item[D.] Privacy Review: Conduct a review of the dataset creation process to address potential privacy concerns and ensure that the data collection respects the rights and consent of the individuals involved.
\item[E.] Model Evaluation Metrics: Design evaluation metrics to assess the performance of NLP models more accurately across diverse linguistic and cultural backgrounds.
\item[F.] Other
\end{enumerate}
\item {[SQ-9b]} To what degree would you consider it an important responsibility of data science teams to develop systems that are accurately representative of users’ experiences through data? * (Ranking: 1 to 5)
\end{enumerate}

\textbf{Diversity in the Data Production Pipeline.} 
\begin{enumerate}
\item {[SQ-10a]} Have you ever considered involving diverse community members with varying demographic backgrounds (gender, age, race/ethnicity, geo-location, etc.), and at what stage of an NLP project pipeline could external community input be useful?
\item {[SQ-10b]} Do you think the demographic background (gender, age, race, nationality, etc.) of data technologists (workers, annotators, engineers, researchers) can affect the outcome of NLP datasets? 
\begin{enumerate}
\item[A.] Yes
\item[B.] No
\end{enumerate}
\item {[SQ-10c]} How feasible might it be to involve external communities of annotators in your NLP data annotation processes and workflows?
\begin{enumerate}
\item[A.] Very feasible 
\item[B.] Feasible
\item[C.] Fairly feasible
\item[D.] Less feasible
\item[E.] Not feasible
\end{enumerate} 
\item {[SQ-10d]} How feasible might it be to involve external communities of annotators in your NLP data annotation processes and workflows?
\begin{enumerate}
\item[A.] Consensus deliberation
\item[B.] Length of time to productively annotate
\item[C.] Understanding metrics of success
\item[D.] Access to data sources
\item[E.] Annotation quality
\item[F.] Understanding the application or technology in which data is used for
\item[G.] I'm unsure
\item[H.] Other
\end{enumerate} 
\end{enumerate}

\textbf{U.S. White House Blueprint for an AI Bill of Rights.}
\begin{enumerate}
\item {[SQ-11a]} Before this survey, how familiar were you with the Algorithmic Discrimination Protections outlined in the U.S. White House Blueprint for an AI Bill of Rights? *
\begin{enumerate}
\item[A.] Not familiar at all
\item[B.] Slightly familiar
\item[C.] Moderately familiar
\item[D.] Very familiar
\item[E.] Extremely familiar
\end{enumerate}
\item {[SQ-11b]} In your opinion, do you believe that the AI Bill of Rights is necessary to ensure responsibility and representation in NLP data-centric systems, safeguarding against algorithmic discrimination and bias? * (Ranking: 1-5 scale)
\item {[SQ-11c]} Review the outlined expectations for Algorithmic Discrimination Protections on pg. 26 of the Blueprint for AI Bill of Rights document, please rate how you perceive the relevance of this expectation on your work. (Multi-select)
(pg. 26) - "The expectations for the automated systems are meant to serve as a blueprint for the development of additional technical standards and practices that are tailored for particular sectors and contexts."
\begin{enumerate}
\item[A.] Proactive assessment of equity in design 
\item[B.] Representative and robust data 
\item[C.] Guarding against proxies 
\item[D.] Ensuring accessibility during design, development, and deployment 
\item[E.] Disparity assessment
\item[F.] Disparity mitigation 
\item[G.] Ongoing monitoring and mitigation
\item[H.] Independent evaluation 
\item[I.] Reporting
\end{enumerate} 
\item {[SQ-11d]} Who should be responsible for ensuring that there are safe-guards in place to prevent AI from being discriminatory, harmful, or unfair? *
Assign a rank from 1 to 6, where 1 indicates the most responsible and 6 indicates the least responsible:
\begin{enumerate}
\item[A.] Users and the General Public
\item[B.] Public Interest Technology Groups (non-profits, non-government agencies, community organizations)
\item[C.] Data Scientists/Engineers
\item[D.] Tech Companies or AI Development Firms
\item[E.] Local Municipalities and State Governments
\item[F.] Federal Policy Makers and Agencies
\end{enumerate} 
\item {[SQ-11e]} Final Thoughts:
Please share any additional comments or insights you have regarding the Algorithmic Discrimination Protections. and their practical applicability in NLP data-centric systems.
\end{enumerate}
\textbf{Demographics.}
\begin{enumerate}
\item Please share your gender identity.
\begin{enumerate}
\item[A.] Male
\item[B.] Female
\item[C.] Non-Binary/Non-Comforming
\item[D.] Prefer not to answer
\item[E.] Other
\end{enumerate}
\item Which race or ethnicity best describes you?
\begin{enumerate}
\item[A.] Black or African American
\item[B.] White / Caucasian
\item[C.] Asian: Far East, Southeast Asia, or the Indian subcontinent
\item[D.] American Indian or Alaskan Native
\item[E.] Hispanic or Latin Origin
\item[F.] Native Hawaiian or Pacific Islander
\item[J.] Prefer not to answer
\item[H.] Other
\end{enumerate}
\item What languages do you speak? Start with native language(s), followed by others.
\item Please include your state of residence.
\item Please indicate your highest educational degree obtained.
\end{enumerate}

\subsection{B. Focus Group Questions}
\begin{enumerate}
\item {[FG-6a]} The domain area with the least number of people having experience is: Mitigating bias in data training sets (~6\%). Why might this be? Why might it be a challenge for NLP practitioners to be less experienced in mitigating bias? 
\item {[FG-6c]} A portion of people mentioned LLMs: How might the introduction of sophisticated LLMs change the ways in which NLP practitioners produce datasets?
\item {[FG-7a]} Some (~21) survey respondents mentioned actively seek diverse sources, regularly reviews of datasets, and incorporating inclusive sampling methods for balanced representation.
- How might NLP practitioners source diverse datasets? What limitations or challenges could make sourcing this data difficult or NLP practitioners?
\item {[FG-7b]} What specific social ethnic, cultural, or language groups may come to mind when considering how NLP data is collected? This may include (minority language communities: African American English speakers; Chicano-American English speakers; American Signed Language speakers; Southern US English speakers; American English speakers with impairments, etc.). 
\item {[FG-7c]} How might the ways in which NLP practitioners are able to account for these characteristics vary based on the type of NLP technology? Are there types of technologies where this is less likely to be a factor - why?
\item {[FG-8a]} How is quality and accuracy defined for data annotators? Who develops metrics of success for data annotators? 
- What tensions might that cause for data annotators? 
- Are you familiar with the concept of “grounded-truth” metrics of success in NLP user-experience research? What assumptions or ideals might users have about how NLP technologies should perform that present tensions in NLP development?
\item {[FG-8b]}What might some of these conflicts be? Does it impede on how annotations can be done?
\item {[FG-9a]} Why might model evaluation metrics be a less prioritized factor? What are current challenges to evaluating NLP system performance and how might this play a role in understanding how well systems work for diverse users? 
\item {[FG-9b]} Why might some NLP practitioners consider this of lesser importance for some NLP technologies? 
\item {[FG-10a]} Can you further discuss what forms of input from diverse community members would be beneficial for these stages? What could you NLP practitioners learn from these external partners?
\item {[FG-10b]} ~96\% of survey respondents thought that the demographic background (gender, age, race, nationality, etc.) of data technologists  can affect the outcome of NLP datasets.
- In what ways could data technologists identities affect the outcome or performance of NLP datasets or systems?
- What are your thoughts on counter-arguments that data technologists identities would not affect NLP system outcomes?
\item {[FG-10c]} An overwhelming number of survey respondents indicated that it was fairly feasible to very feasible to involve external communities of data annotators.
- Why might this be beneficial to NLP technology teams and external communities of data annotators? 
- How might this feasibility vary in practice depending on factors such as the organization policies, intellectual property, and even data privacy? 
- When should diverse communities of external data annotators not be involved? How might NLP practitioners and teams recognize this?
\item {[FG-10d]} “They are experts too and I worry that researchers would not see this.” “Being glossed over or dismissed, particularly as annotators.”
- Why might external community members knowledge and skills be undermined in NLP data annotation work? Who are the actors that influence their potential involvement within private companies and within academic or research-based institutions? 
- How might challenges differ based on the production of commercial products/services and research of NLP dataset production within academic or research-based institutions? 
\item {[FG-11a]} How would NLP practitioners become aware of emerging AI policy and regulations within the U.S.? Whose responsibility is it to inform or education data technologists? 
\item {[FG-11b]} Why might AI Bill of Rights and other forms of AI regulation policy be important to technologists and also the public (users, consumers, customers, residents)? 
\item {[FG-11c]} Can you discuss why these expectations might be ranked in this order? In particular, the top four were among the most relevant to NLP practitioners.
\item {[FG-11d]} - How would you describe responsibility preventing adverse AI impact as a technologist?
-  Can you discuss why these stakeholders might be ranked in this order? Why would Federal, State, and Local governments be the most responsible? Why would Tech/AI Companies or Data Scientists/Engineers be least responsible?
\item {[FG-11e]} How might engaging with international and global AI policy influence U.S. AI policies? Are there other countries doing things that the U.S. could learn from? 
\end{enumerate}

\subsection{C. Supplementary Survey Data}
\begin{table}[!ht]
\centering
\resizebox{\columnwidth}{!}{ 
\begin{tabular}{p{4cm}p{3cm}p{6cm}}
 \toprule 
 \textbf{Category} & \textbf{\# of responses} & \textbf{Responses} \\ 
 \midrule 
    Automatic Speech Recognition (ASR) & 8 & Speech to text dictation apps, Automated speech recognition \\
    \midrule
    Virtual Assistants & 12 & Smart Speakers and Home Devices, Virtual assistant \\
    \midrule
    Sentiment Analysis & 7 & Sentiment analysis for customer reviews\\
    \midrule
    Machine Translation & 4 \\
    \midrule
    Text Classification & 2 & Healthcare related text classification, Media article classifier \\
    \midrule
    Language Generation & 2 & Language generation and annotation with human crowdsourced work \\
    \midrule
    Other Technologies & 3 & spam email detection, clinical text processing for race/ethnicity annotations, and question answering systems\\
 \bottomrule 
\end{tabular}
}
\caption{Example of a specific text or speech NLP-enabled projects (technology, feature, tool) that survey respondents (N = 47) have worked on. They were categorized, with the number of respondents associated with each category indicated.}
\label{table5}
\end{table}

\begin{table}[!ht]
    \centering
    \resizebox{\columnwidth}{!}{ 
        \begin{tabular}[t]{ll}
            \toprule
            \textbf{Highest educational degree obtained} & \textbf{\# of respondents} \\
            \midrule
            Bachelors/Undergraduate & 24 \\
            Advanced Degree (PhD, JD, MD, etc.) & 10 \\
            Masters/Professional Degree & 10 \\
            Some college/Associates & 2 \\
            Certificate/Skill Training Program & 1 \\
            \bottomrule
        \end{tabular}
    } 
    \caption{The highest educational degree obtained by the respondents, providing insight into the educational level of the group.}
    \label{table6}
\end{table}

\end{document}